\documentclass[prl,twocolumn,superscriptaddress,showpacs,aps]{revtex4-1}

\usepackage{color}
\usepackage{graphicx}
\usepackage{bm}
\usepackage{amsmath}

\begin{document}

\newcommand{\MgB}{MgB$_2$}
\newcommand{\Tc}{T_{\text{c}}}
\newcommand{\Hci}{H_{\text{c1}}}
\newcommand{\Hcii}{H_{\text{c2}}}
\newcommand{\fq}{\phi_0}
\newcommand{\lamn}{\lambda_n}

\title{Persistence of Metastable Vortex Lattice Domains in {\MgB} in the Presence of Vortex Motion}

\author{C. Rastovski}
\affiliation{Department of Physics, University of Notre Dame, Notre Dame, Indiana 46556, USA}

\author{K. J. Schlesinger}
\altaffiliation{Present address: Department of Physics, University of California, Santa Barbara, California 93106, USA}
\affiliation{Department of Physics, University of Notre Dame, Notre Dame, Indiana 46556, USA}

\author{W. J. Gannon}
\affiliation{Department of Physics and Astronomy, Northwestern University, Evanston, Illinois 60660, USA}

\author{C. D. Dewhurst}
\affiliation{Institut Laue-Langevin, 6 Rue Jules Horowitz, F-38042 Grenoble, France}

\author{L. DeBeer-Schmitt}
\affiliation{Oak Ridge National Laboratory, Oak Ridge, Tennessee 37831, USA}

\author{N. D. Zhigadlo}
\affiliation{Laboratory for Solid State Physics, ETH Zurich, CH-8093 Zurich, Switzerland}

\author{J. Karpinski}
\affiliation{Laboratory for Solid State Physics, ETH Zurich, CH-8093 Zurich, Switzerland}
\affiliation{Institute of Condensed Matter Physics, EPFL, CH-1015 Lausanne Switzerland}

\author{M. R. Eskildsen}
\email{eskildsen@nd.edu}
\affiliation{Department of Physics, University of Notre Dame, Notre Dame, Indiana 46556, USA}

\date{\today}

\begin{abstract}
Recently, extensive vortex lattice metastability was reported in {\MgB} in connection with a second-order rotational phase transition.  However, the mechanism responsible for these well-ordered metastable vortex lattice phases is not well understood. Using small-angle neutron scattering, we studied the vortex lattice in {\MgB} as it was driven from a metastable to the ground state through a series of small changes in the applied magnetic field.
Our results show that metastable vortex lattice domains persist in the presence of substantial vortex motion and directly demonstrate that the metastability is not due to vortex pinning.
Instead, we propose that it is due to the jamming of counterrotated vortex lattice domains which prevents a rotation to the ground state orientation.
\end{abstract}


\maketitle

The study of vortex matter in type-II superconductors is of great interest, both from a fundamental perspective and as an important limiting factor in practical applications.
Recently, we reported the existence of well-ordered metastable (MS) vortex lattice (VL) phases in {\MgB} single crystals~\cite{Das:2012cfa}. The VL of {\MgB} consists of three hexagonal phases separated by second-order rotation transitions.  Cooling across the phase boundaries, it is possible to lock in long-lived, metastable phases. Such robust and previously unobserved metastability raises the question: What mechanism is responsible for the longevity of the metastable states, preventing them from immediately rotating to the ground state (GS)?

While it was previously argued that vortex pinning is an unlikely explanation for the metastability~\cite{Das:2012cfa}, this assertion was based on the generally weak pinning in {\MgB} and the observation of highly ordered metastable VLs. This contrasts the more disordered configurations found in, e.g., YNi$_2$B$_2$C in connection with hysteresis of a reorientation transition~\cite{Levett:2002ba}. It is important to note, however, that the well-ordered metastable VL configurations in {\MgB} were observed for a static configuration following a cooling or heating across the equilibrium phase boundary and, furthermore, that the dismissal of pinning is not rigorous.

In this Letter we report the results of a series of small-angle neutron scattering (SANS) measurements which directly and conclusively rule out vortex pinning as the cause for the metastable VL phases. By preparing a metastable VL and then inducing vortex motion by small changes in the magnetic field, we resolved coexisting metastable and ground state phases and obtained a quantitative measurement of the transition. Specifically we show that metastable VL domains persist in the presence of substantial vortex motion. This is the first direct demonstration of well-ordered, nonequilibrium VL configurations stabilized by a mechanism other than pinning and opens up a new direction for vortex studies.

Our results lend further credibility to the hypothesis suggested by Das \emph{et al.} that the VL domains act as granular entities, jamming against one another and preventing them from rotating to the ground state~\cite{Das:2012cfa}. This is distinct from the jamming of individual vortices observed in materials with more defects~\cite{Shaw:2012cka} or in connection with artificial pinning potentials~\cite{Barabasi:1999ed,Yoshino:2009ey,Reichhardt:2010cy,Karapetrov:2012hp}. Rather, it is analogous to jamming observed in granular systems, which has recently attracted broad interest~\cite{Liu:2010jx}. The VL in {\MgB} may thus serve as an important model system for jamming studies in general.
The SANS experiments were performed on the D11 beam line at the Institut Laue-Langevin (ILL) and on the CG2 General Purpose SANS beam line at the High Flux Isotope Reactor (HFIR) at Oak Ridge National Laboratory.
To achieve diffraction, the sample and magnet were rotated and/or tilted together in order to satisfy the Bragg condition for the VL planes.  To resolve closely located VL Bragg reflections, very tight collimations of the neutron beam were used (D22 $0.03^\circ$; HFIR $0.07^\circ$ FWHM).
Measurements were performed using the same 200~$\mu$g {\MgB} single crystal as in previous SANS experiments \cite{Das:2012cfa,Pal:2006gi}. The sample was grown using isotopically enriched $^{11}$B to decrease neutron absorption \cite{Karpinski:2003wv}, and had a critical temperature $\Tc = 38$~K and upper critical field $H_{c2} = 3.1$~T.  Measurements were performed at 2~K and 0.5~T applied parallel to the $c$ axis.

\begin{figure}
  \includegraphics{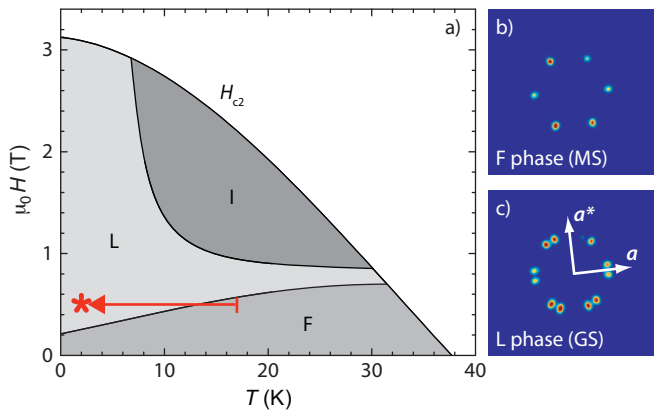}
  \caption{
    {\MgB} vortex lattice phases for $\bm{H} \parallel \bm{c}$. (a) Ground state phase diagram. Diffraction patterns in (b) and (c) were measured at 2~K and 0.5~T, indicated by the red star. The metastable VL (b) is formed by cooling across the {\em F-L} phase boundary as shown by the arrow.  A 50~mT damped field oscillation drives the VL to the ground state configuration (c). Crystalline axes are shown in (c).
    \label{Fig 1}}
\end{figure}

The GS VL phase diagram for {\MgB}, shown in Fig. \ref{Fig 1}(a), consists of three different hexagonal configurations.
In the {\em F} phase, VL Bragg peaks are aligned along the crystal $a$ axis, and in the {\em I} phase along the crystal $a$* axis. In the intermediate {\em L} phase, the VL rotates continuously from the $a$ to $a$* orientation  \cite{Cubitt:2003ip,Das:2012cfa}. Examples of diffraction patterns for the {\em F} and {\em L} VL phases are shown in Figs.~\ref{Fig 1}(b) and \ref{Fig 1}(c), respectively.
The degeneracy in rotation direction of the VL domains in the {\em L} phase results in 12 diffraction peaks, unlike the 6 peaks observed in the {\em F} and {\em I} phases. For a given field and temperature, the GS VL was reached by applying a damped oscillation of the magnetic field with an initial amplitude $\sim$50~mT. A metastable (MS) VL is prepared by heating or cooling across the {\em F-L} or {\em L-I} phase boundaries. The MS VL diffraction pattern in Fig.~\ref{Fig 1}(b) was obtained by preparing a GS VL at 17~K and 0.5~T, and then cooling to 2~K as indicated by the arrow in Fig.~\ref{Fig 1}(a). A field oscillation at 2~K yielded the GS VL diffraction pattern in Fig.~\ref{Fig 1}(c).

\begin{figure}
  \includegraphics{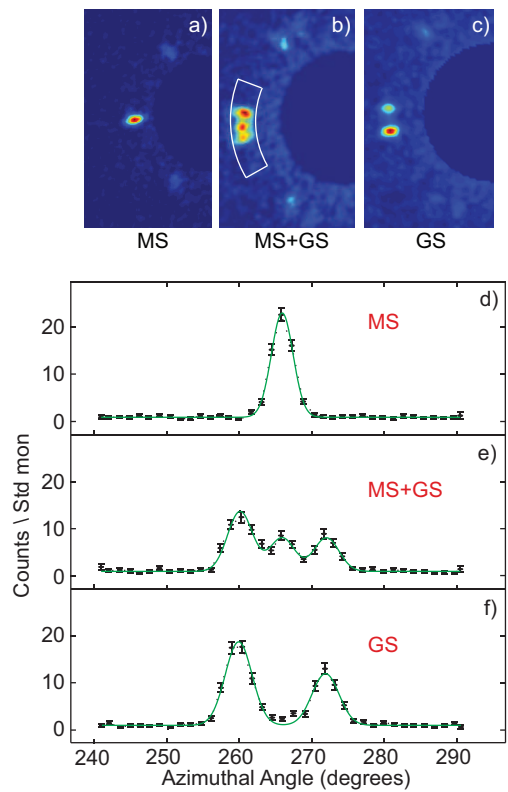}
  \caption{
    Resolving the MS to GS transition. (a)-(c) Diffraction patterns at 2~K: As-prepared MS VL state at 0.5~T (a); coexistence of MS and GS obtained after decreasing the field by 22~mT (b); GS VL obtained after decreasing field by 65~mT (c). (d)-(f) Azimuthal intensity distribution corresponding to the diffraction patterns. Solid lines are fits to the data as described in the text.
    \label{Fig 2}}
\end{figure}

By subjecting the MS VL to small changes in the magnetic field, a gradual transition to the GS is observed. Figure~\ref{Fig 2}(a) shows the as-prepared MS VL at 2~K and 0.5~T.
After reducing the applied magnetic field by 22~mT, VL Bragg peaks corresponding to both the MS and the GS phases are present, as seen in Fig.~\ref{Fig 2}(b). The tight collimation used in the experiment allows a clear resolution of the individual peaks and shows that MS and GS VL domains coexist within the sample. In Fig.~\ref{Fig 2}(c) the magnetic field has been further reduced by a total of 65~mT, driving the VL to the GS within the entire sample.

\begin{figure}
  \includegraphics{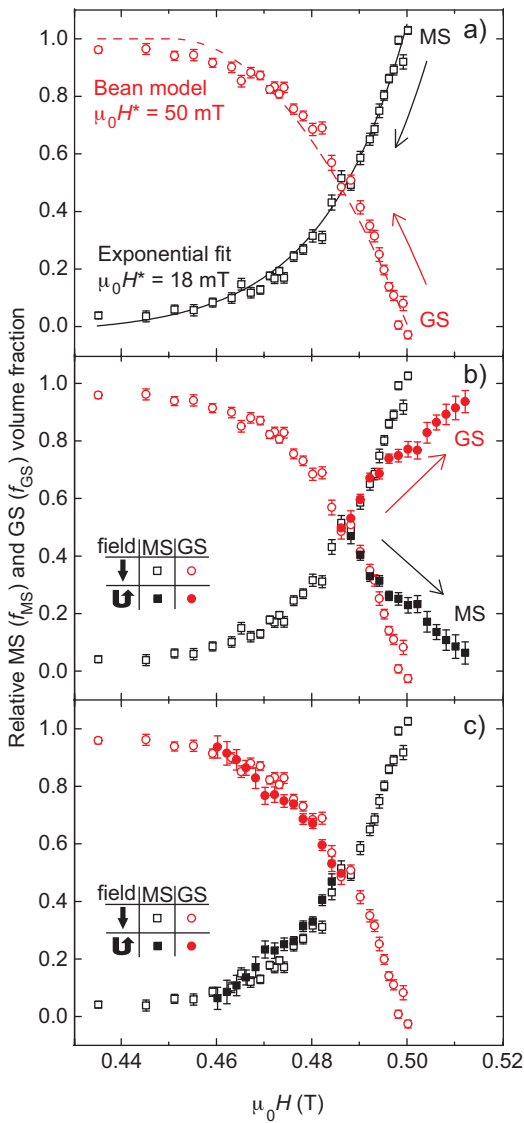}
  \caption{
    VL transition from MS to GS. (a) Relative intensity of the MS and GS VL Bragg peaks measured at 2~K as the applied field was decreased from 0.5~T. The black line is an exponential fit to the data with a characteristic field $H^*= 18$~mT, and the red line is a  Bean model fit with a critical field $H^*=50$~mT. (b) Solid symbols correspond to an initial decrease of 14~mT followed by an increase of the applied magnetic field. Open symbols same as in (a). (c) Same as (b) but with solid symbols reflected about the reversal field of 0.486~T.
    \label{Fig 3}}
\end{figure}

Measurements of the relative intensity of the diffraction peaks makes it possible to determine the fraction of the VL in the GS and MS states. Figures~\ref{Fig 2}(d)-\ref{Fig 2}(f)  show the azimuthal intensity distribution corresponding to Figs.~\ref{Fig 2}(a)-\ref{Fig 2}(c). The data were fitted using three Gaussians  with identical widths and peak centers fixed at $259.9^\circ$, $265.8^\circ$, and $271.8^\circ$, respectively.
The fitted areas under the three peaks ($A_{\text{GS1}}$, $A_{\text{MS}}$, $A_{\text{GS2}}$) provide a quantitative measure of the population of MS and GS VL phases in the sample and allow the calculation of the relative volume fractions:
\begin{eqnarray}
  f_{\text{MS}}&=&A_{\text{MS}}\,/\,(A_{\text{GS1}}+A_{\text{MS}}+A_{\text{GS2}}) \label{fms} \\
  f_{\text{GS}}&=&1-f_{\text{MS}}. \label{fgs}
\end{eqnarray}
Figure~\ref{Fig 3}(a) shows the evolution of the MS and GS VL volume fraction as the vortex lattice is driven from the metastable to the ground state.
Initially, a MS VL was prepared at 0.5~T and 2~K, followed by decreases in the applied field in steps of 2 to 5~mT. The VL was remeasured after every change of the magnetic field,
and $f_{\text{MS}}$ and $f_{\text{GS}}$ were calculated using Eqs.~(\ref{fms}) and (\ref{fgs}). The metastable VL volume fraction smoothly decreased as the applied field was decreased.

A straightforward explanation for VL metastability is that it may be due to vortex pinning. In this picture the VL transition to the ground state is due to vortex motion induced by the decreasing field, as the front of vortex motion propagates from the edges of the crystal toward the center. The red line in Fig.~\ref {Fig 3}(a) is calculated using a simple two-dimensional Bean pinning model with a characteristic field $\mu_0 H^*= 50$~mT \cite{Bean:1964vz}.
Although this provides an overall good fit, there are noticeable deviations in the field ranges $0.43 - 0.46$~T and $0.47 - 0.49$~T. In contrast, a simple exponential function with a characteristic field of 18~mT (black line) provides a better fit throughout the entire field range. In addition, the characteristic field predicted by the Bean model $\mu_0 H^* = \mu_0 J_c D/2$, where $J_c \sim 10^6$~A/m$^2$ \cite{Zehetmayer:2002il, Eisterer:2005jf} and the sample diameter $D \sim 1$~mm, yields a $\mu_0 H^* \sim 1$~mT, a value more than an order of magnitude smaller than that suggested by the data in Fig.~\ref{Fig 3}(a).

To further investigate whether the VL metastability could be due to vortex pinning, a second sequence of measurements was performed using a field reversal. Within the Bean picture, a field reversal will induce a second, inward-moving, flux flow front, and the further transition to the VL ground state should not occur until this front reaches the metastable portion in the center of the sample. A MS VL was prepared and the applied field decreased by 14~mT, which rendered the VL in a state with $f_{\text{MS}} = 53$\%.  The field was then increased in 2~mT increments, causing vortices to reenter the sample from the edges. The results from the field reversal measurements are shown in Fig.~\ref{Fig 3}(b). Contrary to the Bean model prediction, no plateau was observed in the population of the MS VL. Rather, when the data are reflected about the reversal field of $0.486$~T, they coincides within error bars with the results from the decreasing magnetic field measurements, as shown in Fig.~\ref{Fig 3}(c). This shows that any change in the applied field on the order of a few millitesla (consistent with our estimate of $H^*$) is sufficient to perturb some fraction of the metastable  VL domains and cause a further transition to the ground state.

It is possible that the vortex motion is not well described by the simple Bean model. Magneto-optical measurements on thin films of various superconductors including {\MgB} have shown that, in some cases, vortices enter through dendritic avalanches which extend towards the center of the sample and then gradually fill the entire volume \cite{Yurchenko:2009dr}. In a similar scenario, one could imagine a situation with vortex motion confined to dendrites within which the VL has reoriented to the ground state. However, it is important to note that the observed dendritic instability is associated with the initial flux entry into the thin film, whereas our measurements are carried out with a uniform vortex density throughout the sample.

\begin{figure}
  \includegraphics{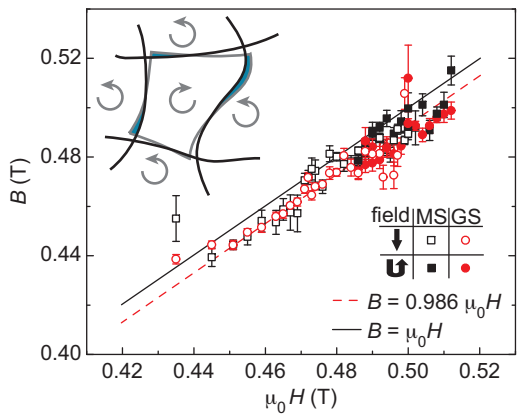}
  \caption{
    Magnetic induction calculated from the measured VL scattering vector as a function of applied magnetic field, corresponding to the data in Fig.~\ref{Fig 3} (b). The dashed line shows a linear fit to the data and the solid line is $B = \mu_0 H$.
    The schematic shown in the inset illustrates the proposed jamming of counterrotated VL domains. For the center domain, rotation to the ground state orientation is prevented, as it would lead to overlapping domains in the shaded regions.
    \label{Fig 4}}
\end{figure}

To definitively resolve whether the VL metastability is due to pinning, we considered the vortex density corresponding to the measurements presented in Fig.~\ref{Fig 3}. For a hexagonal vortex lattice, the VL scattering vector $q$ depends on the magnetic induction $B$ as
\begin{equation}
  q = 2 \pi \sqrt{\frac{2 \, \phi_0}{\sqrt{3} \, B}}.
\end{equation}
The scattering vector, and thus $B$, can be determined experimentally from the peak positions in the VL diffraction patterns. Figure~\ref{Fig 4} shows the magnetic induction as a function of applied field corresponding to the data in Fig.~\ref{Fig 3}(b).
Within the scatter in the data, the magnetic induction for {\em both} the MS and the GS VL domains is found to be proportional to the applied field. The small deviation (1.4\%) of $B/\mu_0 H$ from unity is within the experimental uncertainty. The error on the determination of $B$ exceeds our estimate of $\mu_0 H^* \sim 1$~mT, and a radial peak broadening due to the field variation within the sample is therefore not observed. In contrast, a VL pinned by defects would have a fixed vortex density, independent of changes in the applied field. This unequivocally proves that a mechanism other than pinning is responsible for the VL metastability in {\MgB}.

The absence of pinning, together with the fact that the metastable configurations cannot be understood based on the single domain VL free energy~\cite{Das:2012cfa}, suggests that domain boundaries are responsible for the metastability.  We propose that the metastability is due to a jamming of counterrotated VL domains which prevents a rotation to the ground state orientation, as shown schematically in the inset of Fig.~\ref{Fig 4}.
In this scenario the VL domain boundaries, arising from the two degenerate orientations of the ground state ({\em L} phase), must be robust enough to support a jammed state and ensure the longevity of the metastable VL phases. Furthermore, the domain boundaries must persist even as the vortex lattice expands or is compressed as a result of the changing magnetic field.  Theoretical studies of VL domain boundaries have, to date, been limited \cite{Deutsch:2010bi}, and further work is required to determine the feasibility of the proposed VL domain jamming.  It is important to note that the proposed VL domain jamming represents a novel type of collective vortex behavior, distinct from the jamming of individual vortices observed in artificially engineered pinning potentials in vortex ratchets and similar devices \cite{Barabasi:1999ed,Yoshino:2009ey,Reichhardt:2010cy,Karapetrov:2012hp}. Domain jamming would most likely also be reflected in the dynamical properties of the VL. Analogous to the jamming observed in granular materials one might expect the emergence of power law behavior \cite{Kahng:2001km, DAnna:2001cn, Goodrich:2012ck}, and it is possible that the slow relaxation of the VL found in the noncentrosymmetric superconductor Li$_2$Pt$_3$B is a consequence of VL domain jamming \cite{Miclea:2009ea}.

In summary,  we have performed the first detailed study of the metastable VL in {\MgB} while it was gradually driven to the ground state by small decreases in the applied magnetic field. Our measurements show that metastable VL domains persist in the presence of substantial vortex motion and thus provide definitive evidence that the metastability cannot be ascribed to vortex pinning. Instead, we propose that the metastability in {\MgB} results from the jamming of counterrotating VL domains. Further work is required to explore this hypothesis.

We would like to thank K. Gomes, B. Janko, D. Ray, C. J. Olson Reichhardt, C. Reichhardt, and M. Zehetmayer for valuable discussions, and G. Sigmon for assistance with sample alignment. This work was supported by the Department of Energy, Basic Energy Sciences under Award No. DE-FG02-10ER46783.  The Research at Oak Ridge National Laboratory's High Flux Isotope Reactor was sponsored by the Scientific User Facilities Division, Office of Basic Energy Sciences, U. S. Department of Energy.  Work at ETH Zurich was supported by the Swiss National Science Foundation and the National Center of Competence in Research MaNEP (Materials with Novel Electronic Properties).

\bibliographystyle{apsrev4-1}

\end{document}